\documentclass[3p,11pt,fleqn]{elsarticle}

\usepackage{graphicx}
\usepackage[figuresright]{rotating}
\usepackage{bm,amsmath,amssymb}
\usepackage[mathscr]{eucal}
\usepackage{color}

\biboptions{square,comma,numbers,sort&compress}

\long\def\comment#1{ }
\newcommand{\eqn}[1]{Eq.~\eqref{#1}}
\newcommand{\beq}{\begin{equation}}
\newcommand{\eeq}{\end{equation}}
\newcommand{\nn}{\nonumber\\}
\newcommand{\dif}{{\rm d}}
\newcommand{\rmd}{{\rm d}}
\newcommand{\rme}{{\rm e}}
\newcommand{\rmi}{{\rm i}}
\newcommand{\rmP}{{\rm P}}

\newcommand{\rmtr}{{\rm tr}}

\newcommand{\rmI}{{\rm I}}
\newcommand{\rmJ}{{\rm J}}
\newcommand{\del}{\partial}

\newcommand{\order}[1]{\mcal{O}{(#1)}}
\newcommand{\mcal}{\mathcal}

\newcommand{\bk}{\bm{k}}

\newcommand{\bp}{\bm{p}}
\newcommand{\bx}{\bm{x}}
\newcommand{\by}{\bm{y}}
\newcommand{\bu}{\bm{u}}

\newcommand{\bz}{\bm{z}}

\newcommand{\br}{\bm{r}}

\newcommand{\abar}{\bar{\alpha}_s}

\newcommand{\sdla}{{\rm \scriptscriptstyle DLA}}
\newcommand{\bfkl}{{\rm \scriptscriptstyle BFKL}}
\newcommand{\nlo}{{\rm \scriptscriptstyle NLO}}
\newcommand{\calA}{\mathcal{A}}

\begin{document}

\begin{frontmatter}

\title{{\bf Resumming double logarithms in the QCD evolution \\ of color dipoles}}

\author[sac]{E.~Iancu\corref{cor1}}
\ead{edmond.iancu@cea.fr}

\author[sac]{J.D.~Madrigal}
\ead{jose-daniel.madrigal-martinez@cea.fr}

\author[col]{A.H.~Mueller}
\ead{amh@phys.columbia.ed}

\author[sac]{G.~Soyez}
\ead{gregory.soyez@cea.fr}

\author[ect]{D.N.~Triantafyllopoulos}
\ead{trianta@ectstar.eu}

\address[sac]{Institut de Physique Th\'{e}orique, CEA Saclay, UMR 3681, F-91191 Gif-sur-Yvette, France}
\address[col]{Department of Physics, Columbia University, New York, NY 10027, USA}
\address[ect]{European Centre for Theoretical Studies in Nuclear Physics and Related Areas (ECT*)\\and Fondazione Bruno Kessler, Strada delle Tabarelle 286, I-38123 Villazzano (TN), Italy}

\cortext[cor1]{Corresponding author}

\begin{abstract}
The higher-order perturbative corrections, beyond leading logarithmic accuracy,
to the BFKL evolution in QCD at high energy are well known to suffer from a severe
   lack-of-convergence problem, due to radiative corrections enhanced by
   double collinear logarithms.
   Via an explicit calculation of Feynman graphs in light cone
   (time-ordered) perturbation theory, we show that the corrections
   enhanced by double logarithms (either energy-collinear, or double
   collinear) are associated with soft gluon emissions which are
   strictly ordered in lifetime.
   These corrections can be resummed to all orders by solving an evolution
   equation which is non-local in rapidity.
   This equation can be equivalently rewritten in {\em local} form, but with modified
kernel and initial conditions, which resum double collinear logs to all orders.
   We extend this resummation to the
   next-to-leading order BFKL and BK equations.
   The first numerical studies of the collinearly-improved BK equation demonstrate the
   essential role of the resummation in both stabilizing and slowing
   down the evolution.

\end{abstract}


\end{frontmatter}

\section{Introduction}
\label{sect:intro}

It is by now well established that the Balitsky-JIMWLK
hierarchy\footnote{The acronym JIMWLK stands for Jalilian-Marian,
  Iancu, McLerran, Weigert, Leonidov and Kovner.}
\cite{Balitsky:1995ub,JalilianMarian:1997gr,Iancu:2001ad} and its mean
field approximation known as the Balitsky-Kovchegov (BK) equation
\cite{Kovchegov:1999yj} govern the high-energy evolution of scattering
amplitudes in presence of non-linear effects (multiple scattering and
gluon saturation) responsible for unitarization.
Some of the most remarkable recent developments in that context 
refer to the first calculations of the next-to-leading order (NLO)
corrections \cite{Balitsky:2008zza,Balitsky:2013fea,Kovner:2013ona} to
the B-JIMWLK and BK equations.
These new developments parallel and extend previous efforts, towards
the end of nineties, which established the NLO version
\cite{Fadin:1998py,Ciafaloni:1998gs} of the
Balitsky-Fadin-Kuraev-Lipatov (BFKL) equation
\cite{Lipatov:1976zz,Kuraev:1977fs,Balitsky:1978ic} --- the linearized
version of the BK equation which applies so long as the scattering is
weak.
Although the BFKL and B-JIMWLK equations are based on a common
evolution mechanism, they differ in the way how they treat the scattering problem: 
the BFKL equation deals only with single scattering, as appropriate
for a dilute target, whereas the B-JIMWLK hierarchy
includes the interplay between evolution and multiple scatterings. The
former is usually written in transverse {\em momentum} space, as an equation
for the unintegrated gluon distribution, while the latter is formulated
in terms of transverse {\em coordinates} (better suited suited for implementing
the eikonal approximation) and keeps trace of the multiple scattering of the individual 
partons in the projectile --- each of them represented by a Wilson line.
Such differences explain the difficulty to adapt to the NLO B-JIMWLK
evolution the `collinear resummations' originally developed in the
context of NLO BFKL
\cite{Salam:1998tj,Ciafaloni:1999yw,Ciafaloni:2003rd,Vera:2005jt},
which aim at improving the convergence of the perturbative expansion
for the BFKL kernel.

The collinear resummations refer to perturbative corrections, starting at NLO,
which are enhanced by large, single or double, {\em transverse} logarithms.
Without a proper resummation, which, strictly speaking,
goes beyond the order-by-order expansion of the BFKL kernel, these
large logarithms deprive the NLO BFKL formalism of its predictive
power.

There is no reason to expect this lack-of-convergence problem to be attenuated by the non-linear terms in 
the B-JIMWLK equations: indeed, the `collinear' corrections arise from regions in phase-space 
where the scattering is weak and the non-linear effects are negligible. This was anticipated in 
a semi-analytic study \cite{Triantafyllopoulos:2002nz} and later on confirmed by the
numerical observation that adding a unitarity constraint (in the form of a `saturation boundary')
to the NLO BFKL equation does not help improving the stability of the solution \cite{Avsar:2011ds}.
Very recently, while our work was being completed,  this has been corroborated by a numerical study
 \cite{Lappi:2015fma} of the NLO BK equation \cite{Balitsky:2008zza}: the numerical solution turns
out to be unstable (the scattering amplitude decreases with increasing energy and can even turn
negative) for the physically interesting initial conditions. As also shown in Ref.  \cite{Lappi:2015fma},
this instability can be traced back to a large NLO correction 
to the BFKL kernel enhanced by a double transverse logarithm. This kind of correction, which can be
associated with the choice of the reference scale in the energy logarithm, is well understood at 
BFKL level, where it is successfully resummed to all orders by the schemes proposed in Refs. 
\cite{Salam:1998tj,Ciafaloni:1999yw,Ciafaloni:2003rd,Vera:2005jt}. 
It is our main objective in this paper to propose a similar resummation at the level of the BK equation.

More precisely, our goals are twofold: first, we would like to unambiguously identify the origin
of the double-collinear logarithms in Feynman graphs to all orders and devise a method for
their resummation; second, we would like to reformulate this resummation as a change 
in the kernel of the BK equation, which is energy independent. 
Unlike the corresponding method in the context of NLO BFKL,
where the resummation is generally implemented in double Mellin 
space\footnote{Note however some similarity between our strategy and that proposed in
\cite{Vera:2005jt}, where the $\omega$-shift in Mellin space 
\cite{Salam:1998tj,Ciafaloni:1999yw} has been approximately reformulated as an 
improvement of the BFKL kernel in transverse momentum space.} 
\cite{Salam:1998tj,Ciafaloni:1999yw}, our resummation will
be directly implemented in transverse coordinate space, in order to
be consistent with the non-linear structure of the BK equation.

Concerning the first objective above, our main finding is that the double-collinear
logs arise due to a reduction in the longitudinal phase-space for the high-energy evolution, 
as introduced by the condition that successive gluon emissions be strictly
ordered in lifetime. The interplay between this `kinematical constraint' and the double
transverse logarithms has already been recognized in the literature \cite{Andersson:1995ju,Salam:1998tj}
(see \cite{Beuf:2014uia} for a recent discussion and more references), but we are not aware of any 
systematic derivation of this prescription from Feynman graphs. To emphasize that this is
indeed non-trivial, we notice that double collinear logs are also generated by diagrams with 
{\em anti}-time ordering, but they mutually cancel when all such 
graphs are summed together (see the discussion in Sect.~\ref{sect-DLA} below). 
This observation helps understanding the peculiar way how the double transverse logs arise 
in the context of the NLO BK calculation in \cite{Balitsky:2008zza}.
The main outcome of this diagrammatic analysis is \eqn{finDLA}, which governs the
evolution in the double-logarithmic approximation (DLA): it resums to all orders
the perturbative corrections 
in which each power of the coupling is accompanied by a double logarithm
(either energy-collinear, or double collinear).

\eqn{finDLA} however is  {\em non-local} in `rapidity' (the logarithm of the longitudinal momentum, 
which is our evolution variable), so it does not fully match our goals for a collinearly-improved 
evolution equation\footnote{Collinearly-improved versions of the BK equation which are non-local in 
rapidity have been proposed too
in the literature \cite{Motyka:2009gi,Beuf:2014uia}, but they suffer from some
shortcomings, concerning either the systematics of the resummation (for the approach in 
 \cite{Motyka:2009gi}), or its feasibility in practice (for \cite{Beuf:2014uia}).}.
To cope with that, in Sect.~\ref{sec:resum} we demonstrate that the non-local equation 
\eqref{finDLA} can be reformulated in a {\em local} form, modulo an analytic
continuation and a reshuffling of the perturbative expansion. The new, local, 
equation \eqref{atildeloc} involves an `improved' kernel 
and (for consistency) a modified initial condition, which both resum double-collinear logs to all orders. 

It is then straightforward to extend this resummation to the BFKL and BK equations and thus 
obtain the collinearly-improved BK equation \eqref{resbk}, which is our main result in this paper. 
It is furthermore possible to promote this result to full NLL accuracy, by adding the remaining
NLO BK corrections from Ref.~\cite{Balitsky:2008zza}.
Notice however that the NLO terms include single transverse logarithms, which
may require additional resummations, as was already the case in the context of NLO BFKL
\cite{Salam:1998tj,Ciafaloni:1999yw,Ciafaloni:2003rd}.

Finally, in Sect.~\ref{sec:num} we present the first numerical studies of the resummed
BK equation \eqref{resbk}. These studies clearly demonstrate the role of the resummation in both stabilizing
and significantly slowing down the evolution: the saturation exponent extracted from the numerical
solution is smaller by, roughly, a factor of two than in the absence of the resummation.

\section{The double-logarithmic limit of the BFKL equation}
\label{sect:DLA}

\begin{figure}[t] \centerline{
\includegraphics[width=0.95\textwidth]{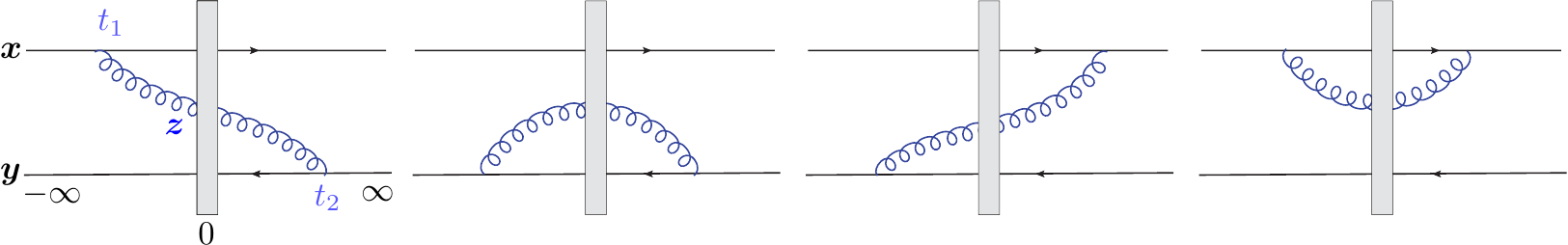}}
 \caption{\sl Diagrams illustrating one step in the BFKL evolution of a $q\bar q$ dipole via the emission
 of a soft `real' gluon (the gluon fluctuation lives at the time of scattering). The target is represented
 as a shockwave. In the (eight) corresponding
 `virtual' graphs, the gluon line is not crossing the shockwave.}
 \label{1gluon}
\end{figure}

In order to fix the notations and for comparison with the more refined
results that we shall later obtain, it is instructive to recall the 
derivation of the `na{\"i}ve DLA', by which we mean the version of
this approximation which neglects the time-ordering of successive
gluon emissions, from the leading-order (LO) BFKL equation
\cite{Lipatov:1976zz,Kuraev:1977fs,Balitsky:1978ic}.
The LO BFKL equation resums the perturbative corrections in which each
power of the QCD coupling $\abar\equiv\alpha_s N_c/\pi$, assumed to be
fixed and small, is accompanied by the energy
logarithm $Y\equiv \ln(s/Q_0^2)$ (the `rapidity'), with $s$ the
center-of-mass energy squared and $Q_0$ a characteristic transverse
momentum scale introduced by the target. In this high-energy
leading-log approximation (LLA), valid when $\abar Y \gtrsim 1$, it is
consistent to treat the scattering and the evolution in the eikonal
approximation.
The LO BFKL equation can then be written as the linearized version of
the BK equation \cite{Balitsky:1995ub,Kovchegov:1999yj}, {\it i.e.} as
an equation for the high-energy evolution of the scattering amplitude
$T_{\bx\by}(Y)$ of a quark-antiquark dipole, with a quark leg at
transverse coordinate $\bx$ and an antiquark leg at transverse
coordinate $\by$, which undergoes weak scattering off a generic target
(a nucleus, or a `shockwave'):
\begin{align}\label{BFKL}
 \frac{\del T_{\bx\by}(Y) }{\del Y}=
 \frac{\abar}{2\pi}\, \int \rmd^2\bz\,
 \mcal{M}_{\bx\by\bz}
 \big[T_{\bx\bz}(Y)+ T_{\bz\by}(Y) -T_{\bx\by}(Y) 
 \big]\,.
\end{align}
This equation involves the  `dipole' version of the BFKL kernel,
 \begin{align}\label{Mdef}
 \mcal{M}_{\bx\by\bz}\,\equiv\, 
 \frac{(\bm{x}-\bm{y})^2}
 {(\bm{x}-\bm{z})^2(\bm{z}-\bm{y})^2}\,,\end{align} which describes the
emission of a soft gluon with transverse coordinate $\bz$ by either
the quark or the antiquark leg of the dipole, followed by its
reabsorption (see Fig.~\ref{1gluon}).
In the limit of a large number of colors $N_c\to\infty$, the positive
quantity $({\abar}/{2\pi}) \mcal{M}_{\bx\by\bz} \rmd^2\bz$ can be
interpreted \cite{Mueller:1993rr} as the differential probability for
the splitting of the original color dipole $(\bx,\by)$ into a pair of
dipoles $(\bx,\bz)$ and $(\bz,\by)$.  The first two terms within the
square brackets, $T_{\bx\bz}$ and $T_{\bz\by}$, are the `real' terms
describing the scattering of the daughter dipoles, whereas the last
one, $-T_{\bx\by}$, is the `virtual' term expressing the reduction in
the probability for the parent dipole to survive at the time of
scattering.

\eqn{BFKL} is valid so long as the scattering is weak, $T\ll 1$,
meaning that all the dipoles look small on the target resolution
scale: $(\bm{x}-\bm{y})^2Q_0^2\ll 1$, etc. 
In this regime, the integration over $\bz$ in the r.h.s. of \eqn{BFKL}
becomes logarithmic when the daughter dipoles are much larger than the
original one, {\it i.e.} for $|\bx-\bz|\simeq |\bz-\by|\gg r\equiv
|\bx-\by|$. For such configurations, the fast decreases of the dipole
kernel, $\mcal{M}_{\bx\by\bz} \simeq r^2/(\bm{x}-\bm{z})^4$, is
partially compensated by the rapid increase of the scattering
amplitudes for the daughter dipoles: $T_{\bx\bz}\simeq
T_{\bz\by}\propto (\bm{x}-\bm{z})^2$.
Note that this property, namely that the dipole scattering amplitude
grows, roughly, like the area of the dipole in the transverse plane,
is indeed satisfied by the initial condition at low energy $Y\simeq 0$
and is also preserved by the high energy evolution in the
(double-logarithmic) approximation discussed below\footnote{In the
  full BK equation, for small enough $r^2$, the same argument applies
  to a large phase-space $1/Q_s^2\gg (\bm{x}-\bm{z})^2 \gg
  r^2$, with $Q_s$ the saturation momentum in the target.}.

The same arguments also imply that, in the region where the
$\bz$--integration is logarithmic, the `virtual' term
$T_{\bx\by}\propto r^2$ is much smaller than the `real' ones and can
be neglected. Its only effect will be to cut off the logarithmic
phase-space at small dipole sizes $|\bx-\bz|\sim r$, or $|\by-\bz|\sim
r$. 
Indeed, when $\bz\to\bx$ or $\bz\to\by$, the linear combination in the
r.h.s. of \eqn{BFKL} vanishes due to `real'--vs.--`virtual'
cancellations and to `color transparency' ($T(Y,r)\to 0$ as $r\to 0$).

To write down the double-logarithmic version of the BFKL equation, it
is convenient to factor out the strong $r^2$--dependence of the
amplitude and write $T_{\bx\by}(Y)\equiv r^2Q_0^2\calA_{\bx\by}(Y)$. 
Focusing on the logarithmically-enhanced contributions only, we can
then average over azimuthal angles and impact parameters, in order to
replace $\calA_{\bx\by}(Y)\to \calA(Y,r^2)$. The ensuing equation for
$\calA(Y,r^2)$, in the (na\"ive) double logarithmic accuracy, is most
conveniently written in integral form, as
 \beq\label{intDLA}
 { \calA(Y,r^2) } =  \calA(0,r^2) +
 {\abar}\int_0^Y\rmd Y_1 \int_{r^2}^{1/Q_0^2}\frac{\rmd z^2}{z^2}\, \calA(Y_1,z^2)\,,\eeq
where the upper limit $1/Q_0^2$ is 
truly the unitarity limit, i.e. the transverse scale where the scattering
becomes strong: $T(Y,r)\sim 1$ when $r\sim 1/Q_0$.
This integral equation can be solved via iterations: $ \calA=
 \sum_{n=0}^{\infty}\calA^{(n)}$, with $\calA^{(n)}$ of order $\abar^n$.
For instance, for the simple
initial condition  $ \calA(0,r^2)=1$, one finds (with $\rho\equiv\ln 1/{(r^2Q_0^2)}$)
 \beq
 \label{A1}
 \calA^{(1)}(Y,\rho)=\abar Y \rho\,,\quad  \calA^{(2)}(Y,\rho)=\frac{\abar^2 Y^2\rho^2}{4}
 \,,\ \cdots \,, \  \calA^{(n)}(Y,\rho)=\frac{(\abar Y\rho)^n}{(n!)^2}\,, \eeq
whose sum is recognized as a modified Bessel function,  $ \calA(Y,\rho) =\rmI_0(2\sqrt{\abar Y\rho})$.

\comment{ \beq\label{DLA}
 \frac{\del \calA(Y,r^2) }{\del Y}=
 {\abar}\int_{r^2}^{1/Q_0^2}\frac{\rmd z^2}{z^2}\, \calA(Y,z^2)\,,\eeq
where the upper limit $1/Q_0^2$ is truly the unitarity limit, i.e.
the transverse scale where the scattering
becomes strong: $T_Y(r)\sim 1$ when $r\sim 1/Q_0$. 
At the present level, it can be thought as the target size; more generally, and after also including
the non-linear terms responsible for unitarity corrections, this will be replaced by the energy-dependent
saturation scale in the target. }

As obvious from the previous discussion, the `na{\"i}ve DLA' equation \eqref{intDLA} resums terms 
of the type $(\abar Y\rho)^n$ to all orders. 
This is a common limit of the BFKL and DGLAP evolution, which however is
not very useful in practice, since valid only in a very narrow regime
(namely, when 
$\abar Y\rho\gtrsim 1$, but such that $\abar Y\ll 1$ and $\abar\rho\ll 1$).  
In the next section, we shall devise 
a more general double-logarithmic approximation, which resums all the perturbative corrections
where $\abar$ is enhanced by exactly two logarithms --- either $Y\rho$ or $\rho^2$.

\section{From Feynman graphs to the DLA equation}
\label{sect-DLA}

To obtain a more general DLA formalism, we shall perform a diagrammatic calculation, using light-cone 
(time-ordered) perturbation theory, of two successive steps in the high energy evolution
of the dipole scattering amplitude. That is, we shall consider 2-gluon graphs in which one
of the emitted gluons is much softer (in the sense of having a much smaller longitudinal momentum $k^+$) 
than the other one. We use conventions where the dipole projectile is an energetic right mover
and we work in the light-cone gauge $A^+=0$.
The use of `old-fashioned' time-ordered perturbation theory is particularly convenient for
our purposes, in that it allows for an economical and physically transparent classification and 
evaluation of the relevant Feynman graphs in the approximations of interest. 

A rather compact way to organize the calculation has been devised in 
Ref.~\cite{Iancu:2014kga}: successive gluon emissions by the projectile,
which are time-ordered and also strongly ordered in $k^+$, are generated by 
repeatedly applying an evolution  `Hamiltonian' on the $S$--matrix for the high-energy 
scattering between the projectile and the target color field $A^-$. 
This $S$--matrix is built from Wilson lines (one for each parton in the projectile), whose
number keeps increasing in the course of the evolution, due to additional gluon emissions. 
For the original dipole, one has ${T}_{\bx\by}=1-{S}_{\bx\by}$
with
 \begin{align}\label{Sdip} {S}_{\bx\by} =  \frac{1}{N_c} \,
\rmtr\big[ V^{\dagger}_{\bx}{V}_{\by} \big]\,,\quad\mbox{where}\quad
 V^{\dagger}_{\bx} \equiv \rmP \exp\left\{\rmi g \int \dif x^+ \,A^-_a(x^+, {\bx}) t^a\right\}.
\end{align}
The Hamiltonian acts via functional differentiation w.r.t. $A^-$ and describes the
emission of a soft gluon out of any of the preexisting Wilson lines,
followed by its
reabsorption (by either the same or a different Wilson line). 
After all the emissions have been produced by acting with the Hamiltonian, one must 
integrate over all the emission times and average over the background field $A^-$
to construct the physical scattering amplitude. This last step though (the functional 
average over $A^-$) is irrelevant  for the present purposes and will be left
unspecified in what follows.

\begin{figure}[t]
 \centering
\includegraphics[width=0.8\textwidth]{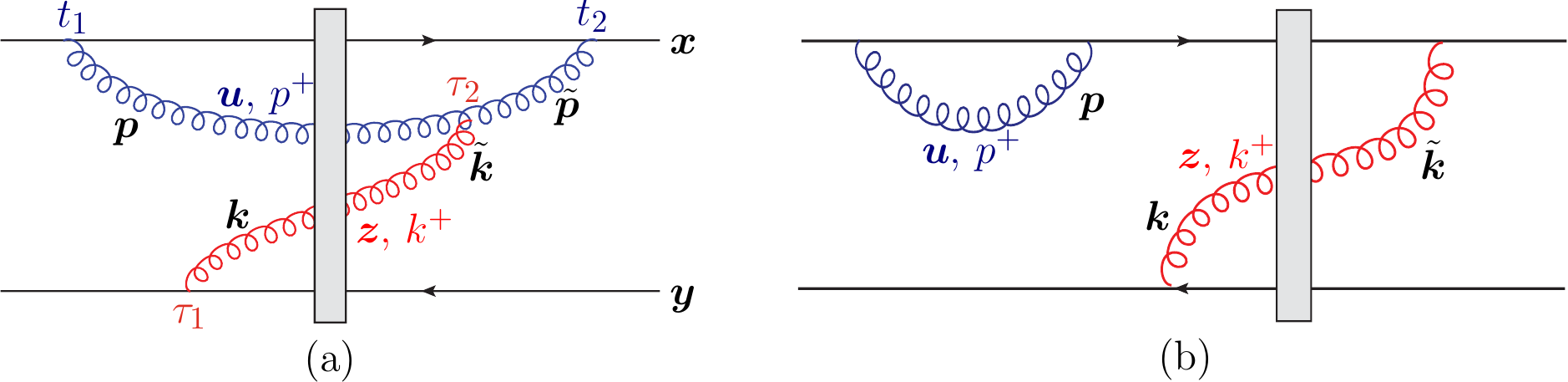}
 \caption{\sl  Diagrams with two gluons which are ordered in longitudinal momentum
 ($p^+>k^+$)
 and also in lifetime ($\tau_p>\tau_k$); (a) a real-real graph; (b) a virtual-real graph.}
 \label{detail}
\end{figure}

\begin{figure}[t]
 \centering
 \includegraphics[scale=0.8]{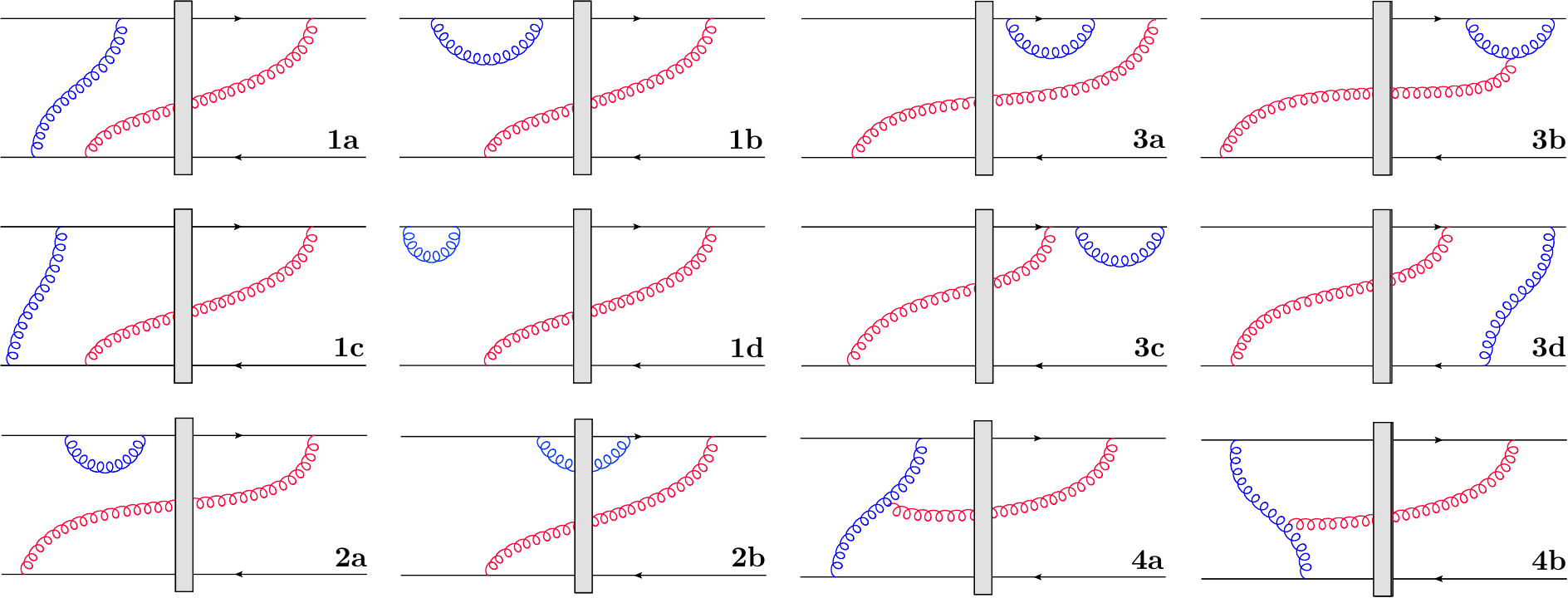}
 \caption{\sl Pattern of cancellations (to DLA) in 2-gluon graphs with
 anti-time-ordering ($\tau_p<\tau_k$).}
 \label{pattern}
\end{figure}

The first two steps in this evolution generate 2--gluon graphs like those shown in Fig.~\ref{detail}, 
where the gluon with longitudinal momentum $p^+$ is emitted first and it is harder than
the gluon $k^+$ ($p^+ > k^+$).  The topologies
(in the sense of time-orderings) selected in Fig.~\ref{detail} are quite special, in
that they contribute already to LLA\footnote{The time-orderings contributing
to LLA can be divided in two classes: \texttt{(i)} if the hard gluon is `real' (it crosses the shockwave),
then it is emitted before the soft one ($t_1< \tau_1$), and reabsorbed after it ($t_2>\tau_2$);
 \texttt{(ii)} if the hard gluon is `virtual', then the 2 gluons have no overlap in time with 
each other --- the hard gluon is emitted and reabsorbed either before the emission of the soft gluon
($t_1 <t_2<\tau_1$) or after the absorption of the latter ($t_2>t_1>\tau_2$).}
: in Fig.~\ref{detail}.a, the hard gluon ($p^+$) is real and is emitted before the soft one ($k^+$), 
but reabsorbed after it; in Fig.~\ref{detail}.b, the hard gluon is virtual and it is both emitted
and reabsorbed prior to the emission of the soft gluon, which is real.
Beyond LLA, other time orderings 
become important as well and will be later considered (see Fig.~\ref{pattern}).

 \comment{\begin{figure}[t] \centerline{\hspace*{-.3cm}
  \includegraphics[width=0.28\textwidth]{BK_dipole_noncross_1.pdf}\quad
\includegraphics[width=0.28\textwidth]{BK_dipole_cross_1.pdf}\quad
\includegraphics[width=0.28\textwidth]{BK_dipole_noncross_2.pdf}}
 \caption{\sl }
 \label{dipole}
\end{figure}
}

We shall first evaluate the 2-real-gluon graph in Fig.~\ref{detail}.a. After integrating over all
emission times, within the ranges $-\infty < t_1 <\tau_1 < 0$ and $0 <\tau_2 < t_2 < \infty$, one
finds the following contribution to the change in dipole $S$-matrix\footnote{To keep
expressions simple, we use the large-$N_c$ limit at intermediate steps,
but some of the final results, notably the DLA equation \eqref{finDLA}, are 
valid for any $N_c$.} (below, $\int_{\bu}\equiv \int {\rmd^2\bu}$ and $\int_{\bp}\equiv \int\frac{\rmd^2\bp}{(2\pi)^2}$)
\begin{align}\label{expr}
-   \frac{g^4N_c^2}{(2\pi)^2} \int_{\bu\bz}S_{\bx\bu}S_{\bu\bz}S_{\bz\by} &
  \int_{\bp\tilde{\bp}\bk\tilde{\bk}}\
    \rme^{\rmi \bp\cdot (\bu-\bx)}\rme^{\rmi \tilde{\bp}\cdot (\bx-\bu)}
  \rme^{\rmi \bk\cdot (\bz-\by)}\rme^{\rmi \tilde{\bk}\cdot (\bu-\bz)}\
\frac{\bp\cdot\tilde{\bp}}{\bp^2 \tilde{\bp}^2}\,\frac{\bk\cdot\tilde{\bk}}{\bk^2\tilde{\bk}^2}\,
  \nn &\qquad\times\int_{q_0^+}^{q^+}\frac{\rmd k^+}{k^+} \int_{k^+}^{q^+}\frac{\rmd p^+}{p^+}\,
  \frac{p^+}{p^++k^+\frac{\bp^2}{\bk^2}}\
  \frac{p^+}{p^++k^+\frac{(\tilde{\bp}-\tilde{\bk})^2}{\tilde{\bk}^2 }}\,.
\end{align}
In the integrals over $k^+$ and $p^+$, the upper limit $q^+$ is the longitudinal momentum
of the quark and antiquark in the original dipole, while the lower limit $q_0^+$ is the longitudinal scale
at which the scattering probes the dipole wavefunction; that is, the overall rapidity interval available
to the evolution of the projectile wavefunction is $Y=\ln(q^+/q_0^+)$.
The denominators in the second line come from time integrations and can be recognized as the usual `energy' 
(here, in the sense of $p^-$) denominators of light-cone perturbation theory. For instance,
\beq
 \frac{p^+}{p^++k^+\frac{\bp^2}{\bk^2}}\,=\,\frac{k^-}{p^-+k^-}\,=\,\frac{\tau_p}{\tau_p+\tau_k}\,,
 \eeq
where $\tau_p\equiv 2p^+/\bp^2 = 1/p^-$ is the lifetime of the hard gluon fluctuation, as determined
by the uncertainty principle, and similarly for $\tau_k$. The integral over $p^+$ is logarithmic provided $p^+$
dominates both energy denominators, that is, so long as\footnote{For the purposes of power counting, 
one can use $|\bk|\sim |\tilde{\bk}|$ and $|\bp|\sim |\tilde{\bp}-\tilde{\bk}|$; indeed, the difference
between e.g. $\bk$ and $\tilde{\bk}$ is due to the scattering off the target, which is a comparatively 
small effect in the high transverse momenta (or small dipole sizes) regime of interest.}
$p^+ > k^+({\bp^2}/{\bk^2})$, or $\tau_p>\tau_k$. 
Hence, to leading logarithmic accuracy for the longitudinal logarithm, 
one can replace ${\tau_p}/({\tau_p+\tau_k})\simeq \Theta (\tau_p-\tau_k)$.

In the BFKL regime, one assumes that there is no strong hierarchy between the transverse momenta,
$|\bk|\sim |\bp|$, so the condition $\tau_p>\tau_k$ is automatically satisfied when $p^+ > k^+$.
In that case, one can freely integrate over transverse momenta in expressions like \eqn{expr}, 
to generate the Weizs\"acker--Williams propagators of the soft gluons,  according to
 \beq\label{WW}\int\frac{\rmd^2\bp}{(2\pi)^2}\,\frac{p^i}
 {\bp^2}\ \rme^{\rmi \bp\cdot(\bx-\bz)}\,=\,-\frac{\rmi}{2\pi}\,
 \frac{x^i-z^i}{(\bx-\bz)^2}\,.\eeq
After also summing over all possible connections for the two emitted gluons, one builds
the relevant product of dipole kernels (i.e., ${\cal M}_{\bx\by\bu}{\cal M}_{\bu\by\bz}$ for the sequence 
of emissions illustrated in Fig.~\ref{detail}).

However, this is strictly correct only so long as the transverse phase-space is by itself not logarithmic,
meaning so long as $Y\gg \rho$, where $\rho\equiv\ln (Q^2/Q_0^2)$ measures the logarithmic
separation in transverse scales between the original dipole, with size $r\equiv 1/Q$, and the target,
with size $1/Q_0$. In the end, the transverse integrations
in \eqn{expr} are restricted to this range, e.g. $Q_0^2\lesssim \bp^2 \lesssim Q^2$ (see below).
For sufficiently large values of $\rho$, one opens the phase-space for a logarithmic integration over
$\bp^2$, which favors relatively large values $|\bp|\gg |\bk|$. In this regime,
the theta-function  $ \Theta (\tau_p-\tau_k)=\Theta(p^+ - k^+ ({\bp^2}/{\bk^2}))$ becomes relevant
and its effect is to reduce the longitudinal phase-space, roughly from $Y$ to $Y-\rho$.

To the accuracy of interest, i.e. to correctly keep both the corrections of orders $\abar Y $ and
$\abar \rho^2$ generated when integrating out the hard gluon $p^+$, the constraint $\tau_p>\tau_k$
can be enforced directly in coordinate space, like $p^+\bar u^2 > k^+\bar z^2$. Here, we have anticipated 
that the corrections of the form $\abar \rho^2$ come from emissions which are strongly ordered in
transverse sizes, such that the daughter dipoles are much larger than the parent one. 
In this regime,
\beq\label{DLAsizes}
|\bz-\bx|\simeq |\bz-\by|\simeq |\bz-\bu|\,\gg\,|\bu-\bx|\simeq |\bu-\by|\,\gg\,
r=|\bx-\by|\,,\eeq
and $\bar u$ refers
to any of the sizes, $|\bu-\bx|$ or $|\bu-\by|$, of the first pair of daughter dipoles, while $\bar z$
similarly refers to the daughter dipoles produced by the second 
splitting\footnote{When integrating over generic values $\bu$ and 
$\bz$, like in \eqn{2real} below,
one can set $\bar u=\mbox{max}\,(|\bu-\bx|,\,|\bu-\by|)$ and $\bar z=\mbox{max}\,(|\bz-\bx|,\,|\bz-\by|,
\,|\bz-\bu|)$.}.
After performing the momentum integrals in \eqn{expr}, summing over
all the possible connections for both emitted gluons, and adding the other splitting sequence
 (where the gluon at $\bz$ is emitted from the dipole $(\bx,\bu)$), one finds the
following result from the 32 {\em time--ordered} graphs
with two `real' gluons (at large $N_c$):
\begin{equation}\label{2real}
 \left(\frac{\abar}{2\pi}\right)^2\int_{q_0^+}^{q^+}\frac{\rmd k^+}{k^+} \int_{k^+}^{q^+}\frac{\rmd p^+}{p^+}\!
  \int_{\bu\bz} \Theta(p^+\bar u^2-k^+\bar z^2){\cal M}_{\bx\by\bu}
 \big [{\cal M}_{\bu\by\bz}S_{\bx\bu}S_{\bu\bz}S_{\bz\by}+{\cal M}_{\bx\bu\bz}S_{\bx\bz}S_{\bz\bu}S_{\bu\by}\big].
\end{equation}
Except for the theta-function enforcing time-ordering, this is recognized as the effect of two consecutive steps in
the LO BFKL evolution. 

To this result, one must add contributions coming from virtual graphs, evaluated to the same accuracy.
The `real-virtual' graphs in which the harder gluon ($p^+$) is virtual, whereas
the softer one ($k^+$) is real, are the only ones that matter to the accuracy of interest. Consider first the 32
such graphs whose topologies (i.e. time-orderings) exist already at LLA, namely those where
the two gluons have no overlap in time with each other (an example is shown in Fig.~\ref{detail}.b).
They give
\begin{equation}\label{virtBFKL}
- \left(\frac{\abar}{2\pi}\right)^2\int_{q_0^+}^{q^+}\frac{\rmd k^+}{k^+} \int_{k^+}^{q^+}\frac{\rmd p^+}{p^+}\,
  \int_{\bu\bz}
{\cal M}_{\bx\by\bu}{\cal M}_{\bx\by\bz}\,S_{\bx\bz}S_{\bz\by}\,.
\end{equation}
In the BFKL context, this contribution is used to regulate the short-distance singularities of \eqn{2real}
as $\bu\to\bx$ and $\bu\to\by$ at a scale set by the original dipole size: $\bar u\gtrsim r$. 
In the present context, it plays a similar role (as anticipated in \eqn{DLAsizes}), except for the fact 
that only the time-ordered piece of \eqref{virtBFKL} is needed for that purpose. That is, albeit
the virtual graphs included in \eqn{virtBFKL} do not naturally involve any time ordering, 
it is nevertheless useful to 
distinguish between the respective time-ordered (TO) and anti-time-ordered (ATO) contributions, 
by inserting $1=\Theta (\tau_p-\tau_k)+\Theta (\tau_k-\tau_p)$ in the integrand of \eqn{virtBFKL}.  
(Here and from now on, $\tau_p=p^+\bar u^2$ and $\tau_k=k^+\bar z^2$.) Then the
TO piece must be combined with the 2-real-gluon contribution in \eqn{2real}, which is itself time-ordered, 
whereas
the ATO piece is to be considered together with other virtual-real graphs, which are naturally ATO
and will be discussed below.

From now on, we shall limit ourselves to the strict double--logarithmic approximation (DLA), where
each power of $\abar$ is accompanied by either $Y\rho$ or $\rho^2$. The corresponding contribution
of \eqn{2real} can be isolated by taking the single scattering approximation and restricting the integrations over $\bu$ and $\bz$ according to \eqn{DLAsizes}. This allows for simplifications like
\beq\label{approx}
{\cal M}_{\bu\by\bz}{\cal M}_{\bx\by\bu}
\simeq \frac{\br^2}{\bar u^2\bar z^4}\,,\qquad 1- S_{\bx\bu}S_{\bu\bz}S_{\bz\by}\,\simeq\,
T_{\bu\bz}+T_{\bz\by}\,\simeq\,2T(\bar z^2)\,.\eeq
For subsequent discussions, it is important to stress that, to DLA, it is only the last emitted gluon 
(the one with the largest tranvese size $\bar z$)  which contributes to scattering.
Then the integrals over $p^+$ and $\bar u$ are both logarithmic, as anticipated, and can be evaluated as
\begin{equation}\label{rho2}
\int^{\bar z^2}_{r^2}\frac{\rmd \bar u^2}{\bar u^2} \int^{q^+}_{k^+\frac{\bar z^2}{\bar u^2}}\frac{\rmd p^+}{p^+}=
\int^{\bar z^2}_{r^2}\frac{\rmd \bar u^2}{\bar u^2}\left(\ln\frac{q^+}{k^+}-\ln\frac{\bar z^2}{\bar u^2}\right)=
Y\rho-\frac{\rho^2}{2}\,,
\end{equation}
where the logarithmic variables $Y=\ln({q^+}/{k^+})$ and $\rho=\ln({\bar z^2}/{r^2})$ refer
to the phase-space available to the hard gluon $p^+$. Note that we have implicitly assumed above
that $Y>\rho$, so that the integral over $p^+$ has indeed support for any $\bar u\ge r$. This
can be recognized
as the condition for the lifetime $\tau_k=k^+\bar z^2$ of the soft gluon fluctuation be (much) smaller
than the `lifetime' $\tau_q=q^+r^2$ of the original dipole (the duration of the quantum process
which has produced that dipole, e.g. the fluctuation of the virtual photon in DIS).

To summarize, by integrating out the intermediate gluon $p^+$, one
has produced, besides the expected LLA contribution $\abar Y\rho$, also a contribution
$\abar\rho^2$, which can be interpreted as a NLO correction to the BFKL kernel for the emission of the soft
gluon $k^+$. This correction matches the respective piece
(that enhanced by a double transverse logarithm) of the full NLO result in Ref.~\cite{Balitsky:2008zza}.
The last remark might suggest that the remaining 2-gluon graphs, that have not been considered so far
and which correspond to other time orderings, do not contribute to order $\abar\rho^2$. But this is not quite
true:
contributions of this order arise from all the diagrams which are {\em anti-time-ordered} (ATO), 
in the sense that the lifetime of the hard gluon is shorter than that of the soft one (to DLA, at least).
Topologically, the class includes two types of diagrams: \texttt{(i)} real-virtual graphs where the 
hard gluon is virtual and overlaps in time with the soft gluon which is real
(some examples are the graphs 1a, 1b, 2a, 3a, 3b, 4a, and 4b in Fig.~\ref{pattern});
\texttt{(ii)} real-real graphs where the hard gluon is emitted after, and absorbed before, the soft
one (see graph 2b in Fig.~\ref{pattern}). To this genuinely ATO diagrams, one must add the ATO pieces
of the virtual-real graphs without overlap in time (see graphs 1c, 1d, 3c, and 3d in Fig.~\ref{pattern}, 
which represent the ATO part of graphs like that in Fig.~\ref{detail}.b, left over from the earlier
calculations),  to cancel UV divergences and introduce an effective
short-distance cutoff equal to $r$ (cf. the discussion after  \eqn{virtBFKL}).

When evaluating graphs of the type \texttt{(i)}  and \texttt{(ii)} above mentioned, one finds that the time 
integrations over the overlapping region produce a factor like
\beq
 \frac{p^-}{p^-+k^-}\,=\,\frac{\tau_k}{\tau_p+\tau_k}\,\simeq\,\Theta(\tau_k-\tau_p)\,,
 \eeq
where the theta-function approximation in the r.h.s. holds in the double-logarithmic region. This
theta-function cuts off the rapidity phase-space at the scale $\rho$ (with $\rho < Y$) and thus
produces a contribution $\propto \rho^2$, as anticipated:
\begin{equation}
\int^{\bar z^2}_{r^2}\frac{\rmd \bar u^2}{\bar u^2} \int_{k^+}^{k^+\frac{\bar z^2}{\bar u^2}}\frac{\rmd p^+}{p^+}=
\int^{\bar z^2}_{r^2}\frac{\rmd \bar u^2}{\bar u^2}\ln\frac{\bar z^2}{\bar u^2}=
\frac{\rho^2}{2}\,.
\end{equation}

It turns out however that all the terms of order $\abar\rho^2$ generated by these ATO graphs exactly cancel
each other. These cancellations can be understood as either the cancellation of `infrared' (large $z^2$) 
logarithms between self-energy and vertex corrections, or as real vs. virtual cancellations for hard gluons
whose scattering is not measured at DLA (cf. the discussion after \eqn{approx}). For instance, the combinations
of graphs 1a and 1b in Fig.~\ref{pattern}, or 3a and 3b, belong to the first category, whereas 2a and 2b
belong to the second. The sum of 1a and 1b leaves an uncompensated UV divergence, which is regulated at the
scale $r$ after also adding the ATO pieces of 1c and 1d (and similarly for 3a, 3b, 3c and 3d). 
Finally, graphs 4a and 4b mutually cancel because the color
current associated with the hard gluon (and responsible for the emission of the soft gluon) has a different
sign in 4a as compared to 4b. Interestingly, the order-$\abar\rho^2$ contribution of the real-real graph 
2b is such that it would exactly compensate the respective contribution of all the {\em time-ordered} real-real 
graphs previously discussed. This explains why, when the calculation is organized in such a way 
that the real-real graphs are all grouped together, as in Ref.~\cite{Balitsky:2008zza} (which
used the standard Feynman rules in momentum space), the net effect of 
order $\abar\rho^2$ is rather seen to arise from the sum of the virtual-real graphs.

The above cancellation pattern for the ATO graphs naturally generalizes to higher orders,
i.e. to graphs involving an arbitrary number of strongly-ordered (in $p^+$) gluon emissions.
This leads us to the main conclusion in this section, namely the fact that the net contributions to DLA come
fully from graphs which, within time-ordered perturbation theory, have exactly the same topology as the graphs 
contributing to LLA, but with the additional constraint 
that the successive gluon emissions must be strictly ordered in lifetimes.
This implies that the perturbative corrections enhanced by the double logarithms $Y\rho$ or $\rho^2$ can
be resummed to all orders by solving a modified version of the DLA equation \eqref{intDLA}, which 
includes the time-ordering condition:
\beq\label{TODLA}
 { \calA(q^+,r^2) } =  \calA(0,r^2) +
 {\abar} \int_{r^2}^{1/Q_0^2}\frac{\rmd z^2}{z^2}\int_{q_0^+}^{q^+\frac{r^2}{z^2}}\frac{\rmd k^+}{k^+}
 \, \calA(k^+,z^2)\,.\eeq
In what follows, we shall mostly use logarithmic variables, with the target scales $q_0^+$ and $Q_0^2$
being the reference scales: $Y=\ln(q^+/q_0^+)$, $Y_1=\ln(k^+/q_0^+)$, $\rho=\ln(1/r^2Q_0^2)$,
$\rho_1=\ln(1/z^2Q_0^2)$. Then \eqn{TODLA} becomes
\beq\label{finDLA}
 { \calA(Y,\rho) } =  \calA(0,\rho) +
 {\abar} \int^{\rho}_0 \rmd \rho_1 \int_{0}^{Y-\rho+\rho_1} \rmd Y_1
 \, \calA(Y_1,\rho_1)\,.\eeq
where a step function $\Theta(Y-\rho+\rho_1)$ is implicitly assumed within the integrand,
to ensure that $Y_1>0$. This integral equation determines the function $\calA(Y,\rho)$ for all 
positive values of $Y$ and $\rho$, but the most interesting physical regime lies at $Y>\rho$.

\section{Resummed kernel for DLA, BFKL, and BK evolutions}
\label{sec:resum}

As compared to its `na\"ive' version in \eqn{intDLA}, the DLA equation with time ordering \eqref{finDLA}
is non-local in rapidity, as it can be best appreciated by rewriting it as a 
differential equation for the $Y$-evolution. This non-locality complicates the practical applications
and, more importantly for our present purposes, it makes it quite tricky to extend
this equation beyond DLA accuracy (albeit similar non-local versions of 
the BFKL and BK equations have been proposed in the literature; see Ref.~\cite{Beuf:2014uia} for
a recent discussion). In what follows, we shall construct an alternative version of this equation,
which is equivalent to \eqref{finDLA}
  in the most interesting physical regime at $Y> \rho$ and which is local in rapidity. The
generalization of this local equation to BFKL and BK will then be almost straightforward, up
to NLL accuracy.

At the mathematical level, it is more convenient to first solve the following problem:
 \beq
 \label{fdla}
 f(Y,\rho) = f(0,\rho)
 + \abar \int_0^{\rho} \dif \rho_1
 \int_{0}^{Y-\rho+\rho_1}
 \dif Y_1 f(Y_1,\rho_1),
 \quad \mbox{with} \quad
 f(0,\rho) = \delta(\rho).
 \eeq
Given its solution $f(Y,\rho) $, one can immediately construct 
the solution to  \eqn{finDLA} for arbitrary initial conditions according to
 \beq
 \label{fdef}
 \mcal{A}(Y,\rho) = \int_0^{\rho} \dif \rho_1\,
 f(Y,\rho-\rho_1) \mcal{A}(0,\rho_1). 
 \eeq
It is straightforward to solve \eqn{fdla} via iterations, to find
 \begin{align}
 \label{fk}
f(Y,\rho)= \delta(\rho)+\sum_{k=1}^{\infty}  f^{(k)}(Y,\rho),\quad\mbox{with}\quad  f^{(k)}(Y,\rho) = \Theta(Y-\rho)\,
 \frac{\abar^k (Y-\rho)^k \rho^{k-1}}{k!(k-1)!}\,,
 \end{align}
where the $\Theta$-function arises from the longitudinal phase-space integration. For the purpose
of the physical interpretation,
it is useful to keep in mind that $f(Y,\rho) $ is essentially the unintegrated gluon distribution in the dipole. 
The presence of the $\Theta$-function in the solution reflects the fact that, in order to emit a soft gluon,
its lifetime $\tau_k=k^+/k_\perp^2$ must be smaller than the coherence time $\tau_q=q^+/Q^2$ of the original dipole.
Summing the above series we arrive at the explicit form
 \beq
 \label{fbess}
 f(Y,\rho) =  \delta(\rho)+\Theta(Y-\rho)
 \sqrt{\frac{\abar(Y-\rho)}{\rho}}\,
 \rmI_1\left(2 \sqrt{\abar(Y-\rho)\rho}\right),
 \eeq
where $\rmI_1$ is the modified Bessel function. Neglecting for the moment the $\Theta$-function one can show that the function above admits an integral representation in the complex plane; namely, one has
$f(Y,\rho) = \Theta(Y-\rho)\tilde{f}(Y,\rho) $, where the new function $\tilde{f}(Y,\rho) $ is defined as
 \beq
 \label{ftildel}
 \tilde{f}(Y,\rho) \equiv
 \int_{\frac{1}{2}- \rmi \infty}^{\frac{1}{2} +\rmi \infty} 
 \frac{\dif \xi}{2\pi \rmi}\,
 \exp\left[\frac{\abar}{1-\xi}\,(Y-\rho) + 
 (1-\xi) \rho\right], 
 \eeq
for any positive $Y$ and $\rho$. This can be viewed as the analytic continuation of the original
function $f(Y,\rho) $  outside the physical range $Y>\rho$.
Making the change of variables $\gamma = \xi + \abar/(1-\xi)$, we can recast \eqn{ftildel} in the form
 of a standard Mellin representation, that is
\beq
 \label{ftildemel}
 \tilde{f}(Y,\rho) =
 \int_{\frac{1}{2}- \rmi \infty}^{\frac{1}{2} +\rmi \infty} 
 \frac{\dif \gamma}{2\pi \rmi}\,
 J(\gamma)
 \exp\left[\abar \chi_\sdla(\gamma) Y + 
 (1-\gamma) \rho\right]. 
 \eeq 
Here, the ``characteristic function'' is determined by
 \beq
 \label{chidla}
 \abar \chi_\sdla(\gamma) = 
 \frac{1}{2}
 \left[
 -(1-\gamma) +
 \sqrt{(1-\gamma)^2 + 4 \abar}  \right] = 
 \frac{\abar}{(1-\gamma)} - \frac{\abar^2}{(1-\gamma)^3}
 + \frac{2 \abar^3}{(1-\gamma)^5} + \cdots,
 \eeq
where all the poles at $\gamma=1$ visible in the r.h.s. are merely 
an artifact of expanding $\chi_\sdla(\gamma)$ in series of $\abar$. The resummed answer is clearly finite at $\gamma=1$. The Jacobian $J(\gamma)$ induced by the change of variables is  related to the characteristic function and reads
 \beq
 J(\gamma) = 1- \abar\,\chi_{\sdla}'(\gamma) = 1 - 
 \frac{\abar}{(1-\gamma)^2} + \frac{3\abar^2}{(1-\gamma)^4}
 + \cdots.
 \eeq
 
The existence of a Mellin representation together with the exponentiation in $Y$ (as manifest in the integrand
in \eqn{ftildemel}) demonstrate that the function $ \tilde{f}(Y,\rho)$ obeys an evolution equation which is
{\em local} in $Y$.  Namely,  \eqn{ftildemel} is tantamount to the following integral equation
 \beq
 \label{ftildeloc}
 \tilde{f}(Y,\rho) = \tilde{f}(0,\rho)
 + \abar \int_0^{Y} \dif Y_1
 \int_0^{\rho}
 \dif \rho_1\, \mcal{K}_\sdla(\rho-\rho_1)
 \tilde{f}(Y_1,\rho_1),
 \eeq
 with the kernel $\mcal{K}_\sdla(\rho)$ defined as the inverse Mellin transform of $\chi_\sdla(\gamma)$, that is\footnote{Such a kernel (albeit written in momentum space) has already appeared in a work focusing on the collinear improvement of the NLO BFKL equation \cite{Vera:2005jt}. In that context, this kernel was obtained as an
approximation to the equation for the  BFKL eigenvalue with $\omega$-shift  \cite{Salam:1998tj}.},
 \beq
 \label{kdla}
 \mcal{K}_\sdla(\rho) = 
 \frac{\rmJ_1\big(2\sqrt{\abar \rho^2}\big)}{\sqrt{\abar \rho^2}}\,=\,1-\frac{\abar \rho^2}{2}+
 \frac{(\abar \rho^2)^2}{12}+\order{(\abar \rho^2)^3}\,,
 \eeq 
($\rmJ_1$ is the Bessel function) and the initial condition $\tilde{f}(0,\rho)$ obtained as the limit of \eqn{ftildemel} at the {\em
unphysical} point $Y=0$~:
\beq
 \label{ftildeic}
 \tilde{f}(0,\rho) = 
 \delta(\rho) - \sqrt{\abar}\, \rmJ_1\big(2 \sqrt{\abar \rho^2}\big).
 \eeq
 
To summarize, the solution to \eqn{ftildeloc} with the kernel \eqref{kdla} and the initial condition 
\eqref{ftildeic} exists for any
positive values $Y$ and $\rho$. For $Y>\rho$ it reduces, by construction,
to the original function $f(Y,\rho)$ in \eqn{fbess}.
The importance of this construction is that it can be immediately generalized to
the evolution of the dipole amplitude, which can be thus rewritten as a local equation in $Y$. 
First, we define the analytic  continuation of $\mcal{A}(Y,\rho)$ according to (cf.~\eqn{fdef})
 \beq
 \label{atildedef}
 \tilde{\mcal{A}}(Y,\rho) \equiv \int_0^{\rho} \dif \rho_1\,
 \tilde{f}(Y,\rho-\rho_1) \mcal{A}(0,\rho_1).
 \eeq
This new function coincides with the physical amplitude $\mcal{A}(Y,\rho)$ for $Y>\rho$. For general,
positive, values of $Y$ and $\rho$, it obeys an equation similar to \eqn{ftildeloc}, that is,
 \beq
 \label{atildeloc}
 \tilde{\mcal{A}}(Y,\rho) = \tilde{\mcal{A}}(0,\rho)
 + \abar \int_0^{Y} \dif Y_1
 \int_0^{\rho}
 \dif \rho_1\, \mcal{K}_\sdla(\rho-\rho_1)
 \tilde{\mcal{A}}(Y_1,\rho_1)\,,
 \eeq
with an initial condition $\tilde{A}(0,\rho)$ which follows from Eqs.~\eqref{atildedef} and
\eqref{ftildeic}. For illustration, consider two interesting initial conditions, namely
$ \mcal{A}(0,\rho) = 1$, which has the advantage of simplicity, and 
$\mcal{A}(0,\rho) = \rho$, which is the limit of the McLerran-Venugopalan (MV) model 
for dipole-nucleus scattering in the single scattering approximation \cite{Iancu:2003xm}. One easily finds
 \begin{align}
 \label{atildeic}
 \tilde{\mcal{A}}(0,\rho) = 
 \begin{cases}
 \displaystyle{\frac{1}{2}} \left [ 1 + 
 \rmJ_0\big(\bar{\rho}\big) \right]
 \quad &\mbox{for} \quad \mcal{A}(0,\rho) = 1,
 \\*[0.2cm]
 \displaystyle{\frac{\rho}{2}}
 \left[1 +  \rmJ_0\big(\bar{\rho}\big)
 +\frac{\pi}{2}\, 
 \rm{\mathbf{H}}_0\big(\bar{\rho}\big)
 \rmJ_1\big(\bar{\rho}\big)
 -\frac{\pi}{2}\, 
 \rm{\mathbf{H}}_1\big(\bar{\rho}\big)
 \rmJ_0\big(\bar{\rho}\big)
 \right]
 \quad &\mbox{for} \quad \mcal{A}(0,\rho) = \rho,
 \end{cases}
 \end{align}
where we have temporarily used the notation $\bar{\rho} =2 \sqrt{\abar \rho^2}$ and where $\rm{\mathbf{H}}_{\alpha}$ is the Struve function.

\eqn{atildeloc} is the sought-after local version of the DLA equation for the dipole amplitude: for $Y>\rho$,
its solution coincides, by construction, with the respective {\em physical} amplitude, i.e. with the 
solution to the non-local equation \eqref{finDLA}. Notice that this rewriting of the DLA evolution in local form is
tantamount to a complete reshuffling of the perturbation series: both the kernel in \eqn{atildeloc} 
and the initial condition in \eqn{atildeic} resum double-collinear
 terms of the type $(\abar\rho^2)^n$ for any $n$. For instance, the very first iteration
of this equation generates all the terms linear in $\abar Y$, i.e. the terms of the type
$\abar Y\rho(\abar\rho^2)^n$ with $n\ge 0$, that would be produced by iterating
the original equation \eqref{finDLA} to all orders. 
Remarkably, even though both the kernel and the initial condition exhibit
oscillations as functions of $\rho$, their combined effect within equations like \eqref{atildeloc} or
\eqref{ftildeloc}  yields a 
solution which is positive definite in the physical region $Y>\rho$, order by order in $\abar$ 
(e.g., this produces the perturbative solution \eqref{fk} for $f(Y,\rho)$).

As we now explain, it is rather straightforward to promote this local
DLA equation into a more complete equation, which includes the right BFKL and BK physics to NLL accuracy.
To that aim, and starting with \eqn{atildeloc}, we shall make 
backwards the steps leading from the LO BFKL equation \eqref{BFKL}
to the `na\"ive' DLA equation \eqref{intDLA}, that is:

 \texttt{(i)} we use the full expression for the dipole scattering amplitude, and more precisely its analytic continuation $\tilde{T}(Y,\rho)\equiv \rme^{-\rho} \tilde{\mcal{A}}(Y,\rho)$
(which coincides with the physical amplitude for $Y>\rho$);

\texttt{(ii)} we return to the use of transverse coordinates in our notations, meaning that we replace $\rho= 
\ln({1}/{r^2 Q_0^2})$, $\rho-\rho_1=\ln({z^2}/{r^2})$, $\tilde{T}(Y,\rho)=\tilde{T}_{\bx\by}(Y)$, and
$2\tilde{T}(Y,z^2)\to \tilde{T}_{\bx\bz}(Y)+\tilde{T}_{\bz\by}(Y)$;

 \texttt{(iii)} we restore the full dipole kernel
by replacing $(r^2/z^4)\rmd z^2\to \mcal{M}_{\bx\by\bz}\rmd^2\bz/\pi$; 


\texttt{(iv)} we reintroduce the
virtual term and at the same time remove the infrared and ultraviolet cutoffs on the integral over $\bz$,
since they are not needed anymore;

\texttt{(v)} we replace the argument of the (additional) kernel $\mcal{K}_\sdla$, that is,
$\rho-\rho_1=\ln({z^2}/{r^2})$, according to $\ln(z^2/r^2) \to  \sqrt{L_{\bx\bz r} L_{\by\bz r}}$, where\footnote{Some 
caution is required here, since there is a small region of integration
where the product appearing under the square root can be negative; in this case, it is enough to let 
$L_{\bx\bz r} L_{\by\bz r}$ $\to |L_{\bx\bz r} L_{\by\bz r}|$ and $\rmJ_1 \to \rmI_1$ (cf. \eqn{kdla}
with $\rho^2 < 0$).}
$L_{\bx\bz r}\equiv \ln[(\bx-\bz)^2/(\bx-\by)^2]$ and $r^2=(\bx-\by)^2$.

The last step above is the only one which is new as compared to the original discussion in Sect.~\ref{sect:DLA}
and will be thoroughly justified in a moment. We are thus led to the following equation 
 \beq
 \label{resbk}
 \frac{\del \tilde{T}_{\bx\by}(Y)}{\del Y} = 
 \frac{\abar}{2\pi}\!
 \int\! \dif^2\bz \mcal{M}_{\bx\by\bz}\,
 \mcal{K}_\sdla\!
 \left(\!\sqrt{L_{\bx\bz r} L_{\by\bz r}}
 \right)
 \!\left[\tilde{T}_{\bx\bz}(Y) \!+\! \tilde{T}_{\bz\by}(Y) \!-\! \tilde{T}_{\bx\by}(Y) \!-\! \tilde{T}_{\bx\bz}(Y) \tilde{T}_{\bz\by}(Y) \right],
 \eeq
where we have also added the non-linear term familiar from the BK equation, to account for multiple
scattering and thus ensure unitarization. \eqn{resbk} improves over the LO BK equation 
by resumming the double-collinear logs, i.e. the perturbative corrections 
of the form $(\abar \rho^2)^n$, to all orders. Importantly, this resummation affects {\em both} the
kernel and the initial conditions. 

So far, the initial condition at $Y=0$ has been specified only in the weak scattering regime where $\tilde{T}\ll 1$
and the precise normalization of $ \tilde{T}(Y=0,r)$ was unessential (since the respective evolution was linear). 
For the purposes of the non-linear equation \eqref{resbk}, however, we need to fix this normalization. 
To that aim, it is convenient to use the MV model \cite{Iancu:2003xm}, which amounts to 
exponentiating the amplitude for a single scattering:
 $ \tilde{T}(0,r)=1-\rme^{-\tilde{T}_0(r)}$, where $\tilde{T}_0(r) =  \rme^{-\rho} \tilde{\mcal{A}}(0,\rho)$, with
  $\rho=\ln({1}/{r^2 Q_0^2})$ and $\tilde{\mcal{A}}(0,\rho)$ as shown in the second line
 in \eqn{atildeic}. 

To fully motivate \eqn{resbk}, we still need to
explain our choice for the argument of $\mcal{K}_\sdla$ in this equation. Clearly, the replacement made at point
\texttt{(v)} above is irrelevant at the present interest (improved LLA), but has two important virtues:
(a) it switches off the effect of the collinear resummation for small daughter dipole sizes,
$|\bx-\bz|\lesssim r$, or $|\by-\bz|\lesssim r$, where this resummation should indeed  play
no role, and (b) when expanded to
second order, i.e. $\mcal{K}_\sdla(\rho)\simeq 1-\abar\rho^2/2$, it precisely
matches the double-logarithmic term contained in the full NLO BK result \cite{Balitsky:2008zza}.
The last feature makes it straightforward to formally
extend \eqn{resbk} to full NLL accuracy: it is sufficient to
add to its r.h.s. all the NLO BK corrections computed in \cite{Balitsky:2008zza}, {\em except} for the
double-log term that has already been included in the kernel. 

Let us finally remind that, strictly speaking, the solution to \eqn{resbk} can be trusted only for
sufficiently small values of $\rho\lesssim Y$ (which in turn requires $Y$ to be large enough,
$\abar Y \gtrsim 1$, in order to significantly evolve away from the `unphysical' initial condition).
In practice though, we expect the BFKL evolution encoded in  \eqn{resbk} to eventually
wash out the oscillations introduced by the initial condition at large $\rho$ and thus progressively built 
a physical tail including at $\rho > Y$. This will be checked via numerical calculations in the next section.

\section{Numerical tests}
\label{sec:num}

In this section we present a brief selection of first numerical studies which illustrate some subtle issues previously 
discussed (like the interplay between local and non-local evolution equations) 
and also some physical consequences of the resummation.

\begin{figure}[t] \centerline{\hspace*{-.3cm}
  \includegraphics[width=0.48\textwidth]{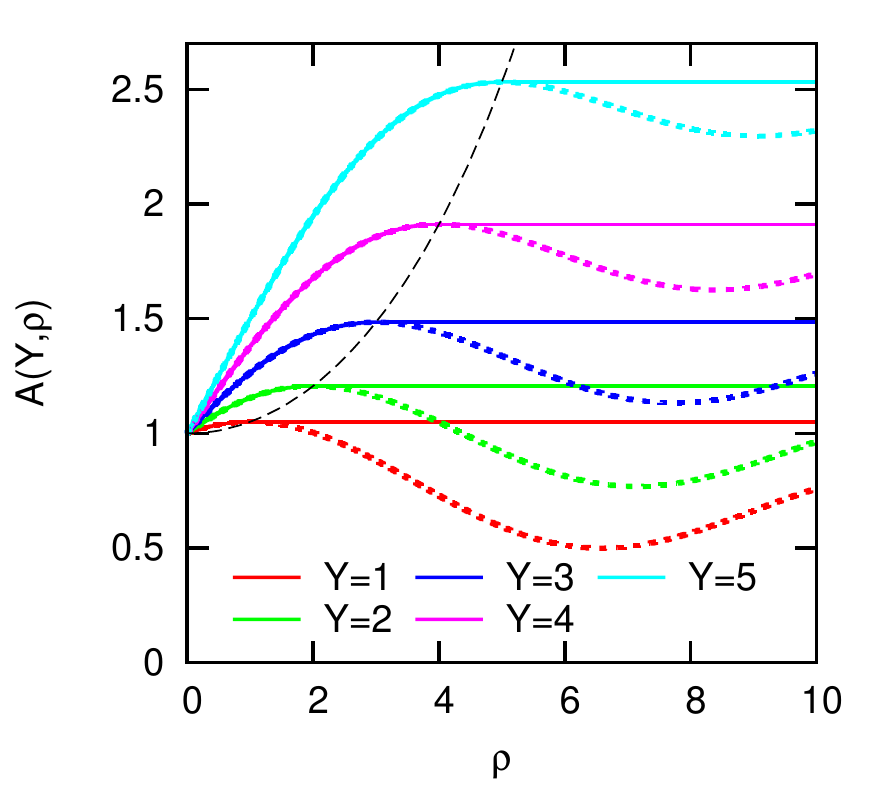}\qquad
\includegraphics[width=0.5\textwidth]{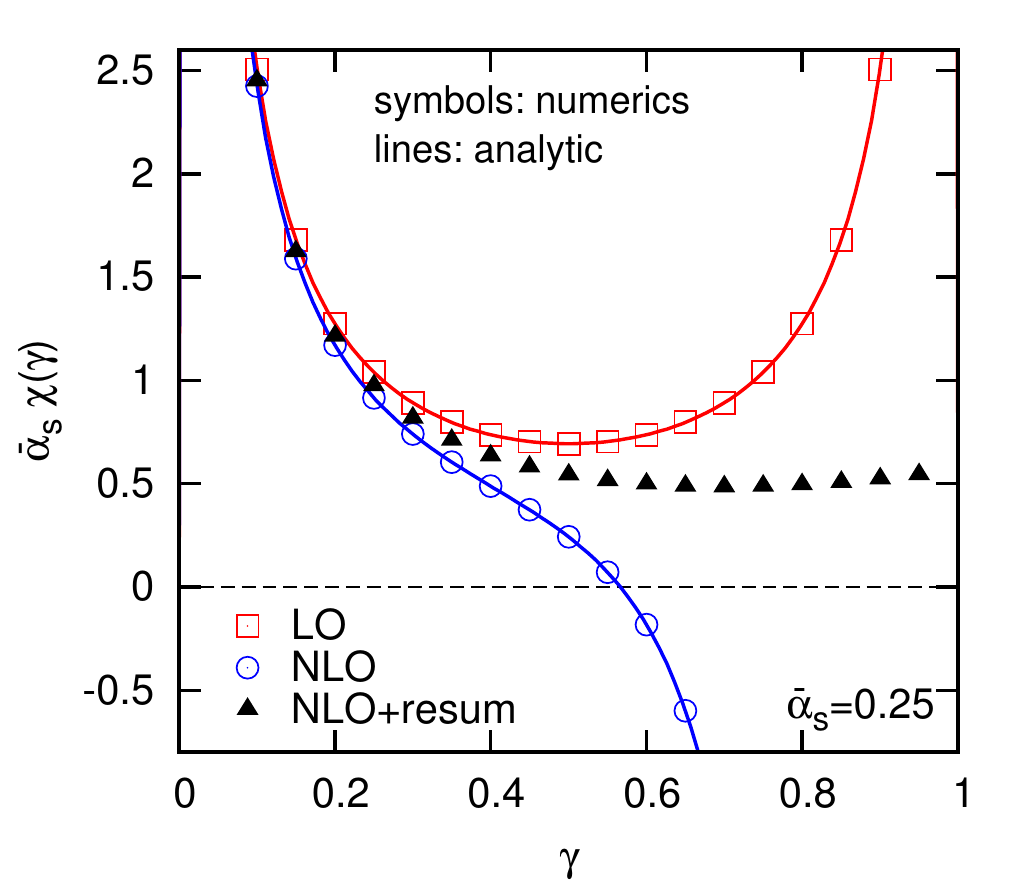}}
 \caption{\sl Left: the DLA solutions for ${\mcal{A}}(Y,\rho)$, cf.  \eqn{finDLA} (full lines),
 and respectively $\tilde{\mcal{A}}(Y,\rho)$, cf.  \eqn{atildeloc} (dashed lines), for various rapidities
 and  (physical) initial
condition ${\mcal{A}}(0,\rho)=1$. Right: the characteristic function $\abar\chi(\gamma)$
as numerically extracted from the kernel in   \eqn{resbk} vs. its LO (BFKL) and `NLO' approximations.}
 \label{ADLA}
\end{figure}

Consider first the double-logarithmic approximation. In the left-hand plot of Fig.~\ref{ADLA}, 
we show the function ${\mcal{A}}(Y,\rho)$, which we recall is related to the dipole amplitude  
${T}(Y,\rho)$, as obtained from two different approaches\footnote{In both cases, we found that a simple
  discretisation of the integral and a Euler method to solve the
  rapidity evolution was sufficient to reach good numerical
  accuracy.}: 
(a) by directly solving the non-local equation \eqref{finDLA} with initial
condition ${\mcal{A}}(0,\rho)=1$, and (b) by solving the local equation \eqref{atildeloc} with the
initial condition shown in the first line of \eqn{atildeic} (which, strictly speaking, truly yields
the analytic continuation $\tilde{\mcal{A}}(Y,\rho)$). The respective solutions are supposed to coincide 
only at $\rho<Y$, where they both represent the actual physical result. This is indeed confirmed by  
the numerical simulations. On top of that, for  $\rho > Y$, we see that the (physical) solution to \eqn{finDLA} 
is independent of 
$\rho$ and equal to ${\mcal{A}}(Y,\rho)=\cosh\sqrt{\abar}\, Y$ \cite{prep}, whereas its analytic continuation
$\tilde{\mcal{A}}(Y,\rho)$ shows (non-physical)
oscillations which are inherited from the initial condition.

We now move to the collinearly-improved BK equation  \eqref{resbk}. 
Numerically, we solve the evolution equation following
a strategy similar to the one described in the Appendix of Ref.~\cite{Alvioli:2012ba}. 
By acting with the kernel in this equation on the power-like test function $r^{2\gamma}$,
one can numerically extract the characteristic function $\abar\chi(\gamma)$ (the would-be Mellin 
transform of the resummed kernel). In the right-hand plot of Fig.~\ref{ADLA}, 
we compare the function $\abar\chi(\gamma)$ 
thus obtained for the particular value $\abar=0.25$ (black triangles)
with the respective LO (BFKL) approximation $\abar\chi_0(\gamma)$ (red squares) 
and with a `NLO' approximation\footnote{In this section, by
  `NLO' we refer to the inclusion of the NLO corrections which are
  enhanced by double collinear logarithms, finite NLO corrections
  being neglected.}, $\abar\chi_\nlo(\gamma)$, obtained by keeping 
only the $\abar$ term in the expansion of $\mcal{K}_\sdla$, that is,
$\mcal{K}_\sdla\to \mcal{K}_\nlo(\rho)\equiv 1-\abar\rho^2/2$.  
The solid lines show the expected analytic results for the LO and NLO
curves and their agreement with the numerical results is a powerful
check that the numerical procedure is under control. As manifest on this figure,
the behavior near $\gamma=1$ is strongly influenced by the higher order corrections.
This can also be understood by inspection of the DLA approximation $\chi_\sdla(\gamma) $ in 
\eqn{chidla}.  At NLO accuracy, $\chi_\sdla(\gamma)$ exhibits a cubic pole at $\gamma=1$,
the second term in the r.h.s. of \eqn{chidla}, with a negative residue
which makes the function $\chi_\nlo(\gamma)$ unstable in the collinear
limit $\gamma\to 1$ (in particular, there is no saddle point on the
real axis). By contrast, the all-order resummation ensures a smooth behavior near
$\gamma=1$, as already noticed after \eqn{chidla}. For $\abar=0.25$,
the function $\chi(\gamma)$ is seen to be almost flat for
$\gamma\gtrsim 0.5$.

 A crude estimate of the saturation line\footnote{We recall
the saturation line $\rho_s(Y)$ is defined by the condition that $T(Y,\rho)\sim 1$ when $\rho=\rho_s(Y)$.}
based on the DLA result in \eqn{fbess} yields \cite{prep}
\beq\label{lambdas}
\rho_s(Y)\equiv \ln\frac{Q_s^2(Y)}{Q_0^2}\simeq
\lambda_s Y\,,\quad\mbox{with}\quad \lambda_s =\frac{4\abar}{1+4\abar}\,,
\eeq
which is significantly smaller than the respective LO result (no resummation) $\lambda_\bfkl
\simeq 4.88\abar$ \cite{Iancu:2003xm}. 
This suggests that the reduction of the longitudinal
phase-space coming from time-ordering and giving rise to 
collinear double logs leads to a considerable reduction in the speed of the evolution.

\begin{figure}[t] \centerline{\hspace*{-.3cm}
  \includegraphics[width=0.35\textwidth]{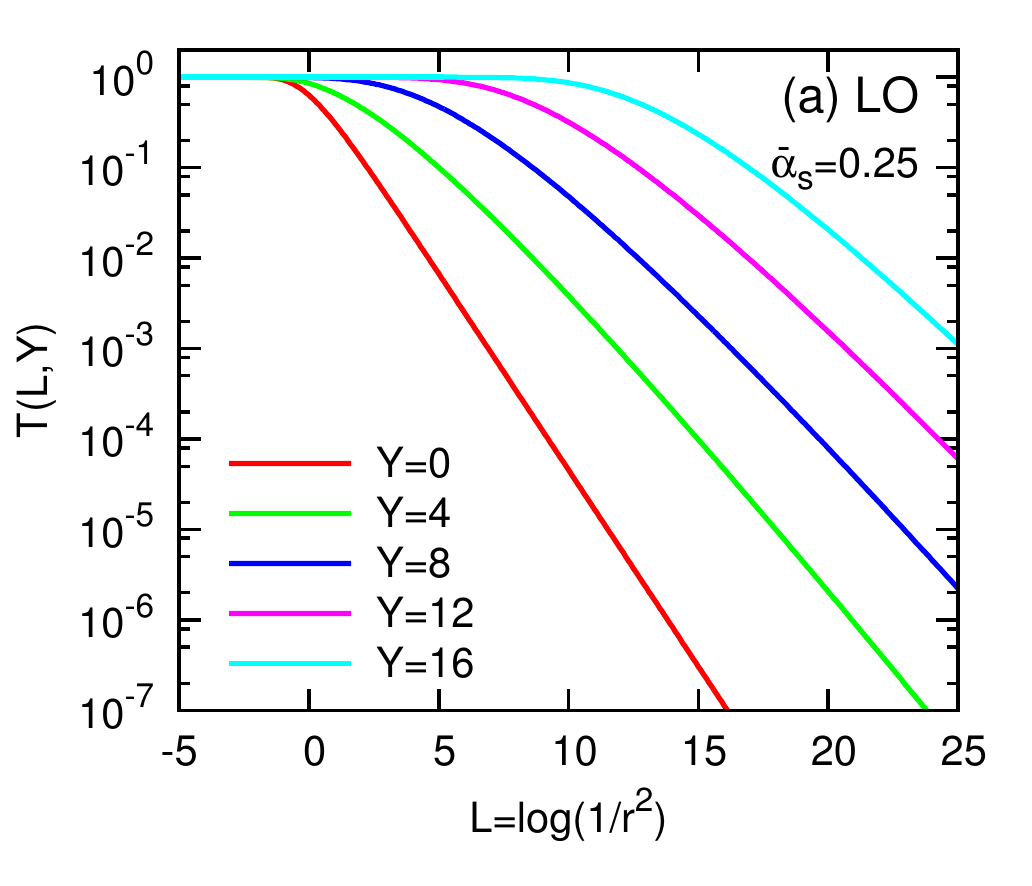}\quad
\includegraphics[width=0.35\textwidth]{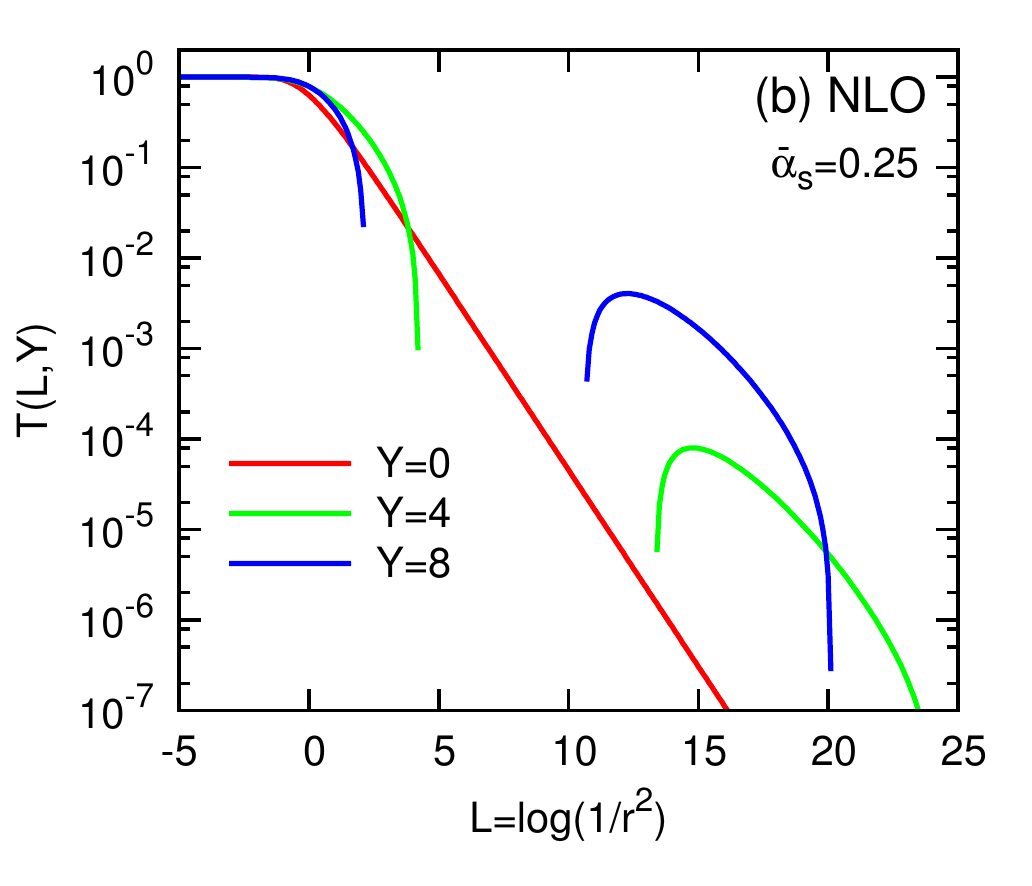}\quad
\includegraphics[width=0.35\textwidth]{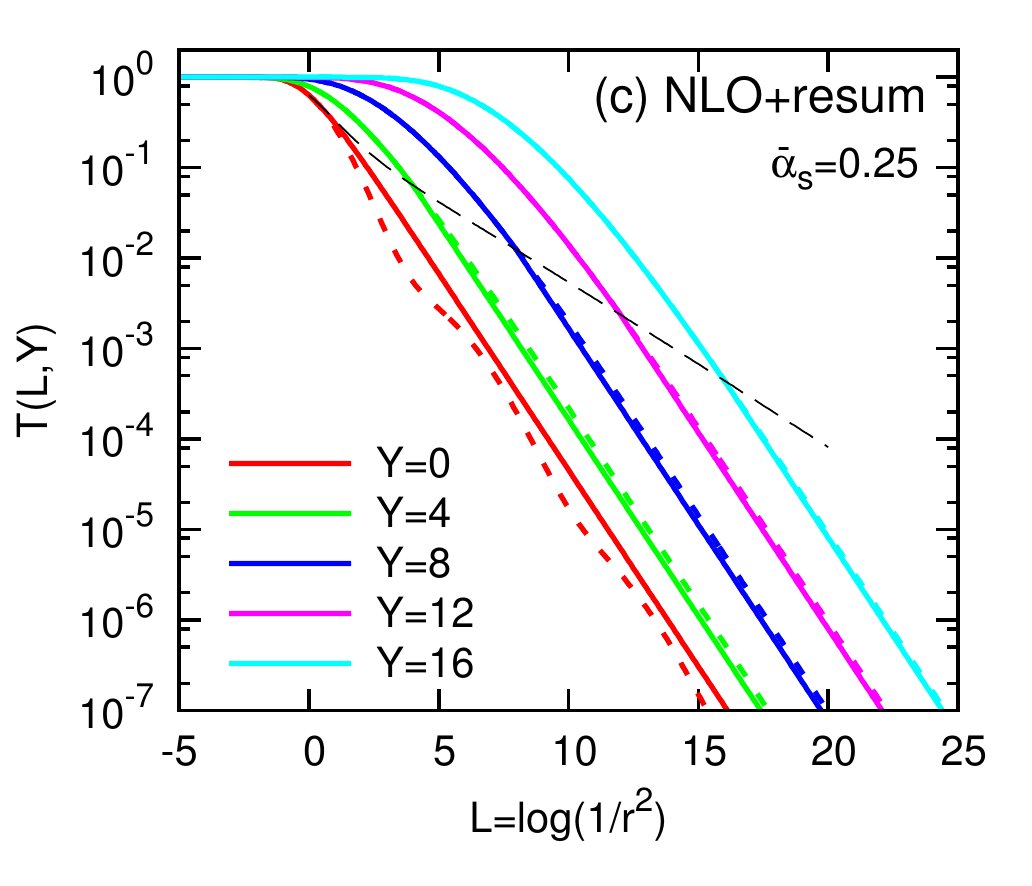}}
 \caption{\sl Numerical solutions to the BK equation for the dipole amplitude at strict LO (i.e.
  \eqn{resbk} with  $\mcal{K}_\sdla\to 1$), NLO (meaning with kernel
$\mcal{K}_\sdla\to \mcal{K}_\nlo$), and
 after resummation (i.e. with the full kernel $\mcal{K}_\sdla$ of \eqn{kdla}).
  The dashed line in fig. (c) indicate the transition between
  $Y<\rho$ and $Y>\rho$; dotted lines are the direct
  result of the numerical simulation, while solid lines have been
  matched to the expected physical behaviour for $\rho>Y$, i.e. $T\propto\rme^{-\rho}$.}
 \label{Tnumerics}
\end{figure}
This expectation is indeed confirmed by the numerical solutions to \eqn{resbk}. In Fig~\ref{Tnumerics}, 
we show the results for $\abar=0.25$ and for an initial condition of the MV type,
with ${\mcal{A}}(0,\rho)=1$ (and hence $\tilde{\mcal{A}}(0,\rho)$ as given in the first line of \eqn{atildeic}).
As before, the results with all-order resummation (cf.
Fig~\ref{Tnumerics}.c) are compared to the respective predictions of LO BFKL (cf.
Fig~\ref{Tnumerics}.a) and to the `NLO' results obtained by using $\mcal{K}_\nlo(\rho)= 1-\abar\rho^2/2$
 (cf. Fig~\ref{Tnumerics}.b). The latter are highly unstable and physically meaningless --- the evolution
 rapidly leads to a {\em negative} scattering amplitude --- as it could have been anticipated in view
 of the pathological behavior of the corresponding characteristic function $\chi_\nlo(\gamma)$ in Fig.~\ref{ADLA}.
 Similar instabilities have been recently observed \cite{Lappi:2015fma} in numerical simulations of
 the full NLO BK equation  and they have been traced back to the large
 double-logarithmic terms $\sim\abar\rho^2$ in the NLO kernel, in agreement with our present findings. 
 By contrast, the evolution with the fully resummed kernel, shown in Fig~\ref{Tnumerics}.c, 
 is perfectly smooth. We also see in Fig~\ref{Tnumerics}.c that the non-physical
oscillations at $\rho>Y$ introduced by resummation in the initial
condition tend to disappear at larger rapidities.
Finally, by comparing the LO results in Fig~\ref{Tnumerics}.a to the resummed ones in Fig~\ref{Tnumerics}.c,
one clearly sees the anticipated reduction in the evolution speed. A quick estimate of the
saturation exponent from the numerical results in
Fig~\ref{Tnumerics}.c yields $\lambda_s\simeq 0.55$, in remarkable
agreement with the crude DLA estimate in \eqn{lambdas}.

\section*{Acknowledgments}
This work is supported by the European Research Council under the Advanced Investigator Grant ERC-AD-267258 and by the Agence Nationale de la Recherche project \# 11-BS04-015-01. The work of A.H.M. is supported in part by the US Department of Energy. Diagrams have been created with Jaxodraw \cite{Binosi:2003yf}.



\begin{thebibliography}{10}

\bibitem{Balitsky:1995ub}
I.~Balitsky,
  \href{http://dx.doi.org/10.1016/0550-3213(95)00638-9}{{\em Nucl. Phys.} {\bf
  B463} (1996)  99--160},
\href{http://arxiv.org/abs/hep-ph/9509348}{{\tt arXiv:hep-ph/9509348}}.

\bibitem{JalilianMarian:1997gr}
J.~Jalilian-Marian, A.~Kovner, A.~Leonidov, and H.~Weigert,
  {\em Phys. Rev.} {\bf D59} (1998) 014014,
   \href{http://xxx.lanl.gov/abs/hep-ph/9706377}{{\tt arXiv:hep-ph/9706377}}.

\bibitem{Iancu:2001ad}
E.~Iancu, A.~Leonidov, and L.D.~McLerran,
  \href{http://dx.doi.org/10.1016/S0370-2693(01)00524-X}{{\em Phys. Lett.} {\bf
  B510} (2001)  133--144},
\href{http://arxiv.org/abs/hep-ph/0102009}{{\tt arXiv:hep-ph/0102009}}.

\bibitem{Kovchegov:1999yj}
Y.V.~Kovchegov,
  \href{http://dx.doi.org/10.1103/PhysRevD.60.034008}{{\em Phys. Rev.} {\bf
  D60} (1999)  034008},
\href{http://arxiv.org/abs/hep-ph/9901281}{{\tt arXiv:hep-ph/9901281}}.

\bibitem{Balitsky:2008zza}
I.~Balitsky and G.A.~Chirilli, 
\href{http://dx.doi.org/10.1103/PhysRevD.77.014019}{{\em
  Phys.Rev.} {\bf D77} (2008)  014019},
\href{http://arxiv.org/abs/0710.4330}{{\tt arXiv:0710.4330 [hep-ph]}}.

\bibitem{Balitsky:2013fea}
I.~Balitsky and G.A.~Chirilli,   
\href{http://dx.doi.org/10.1103/PhysRevD.88.111501}{{\em Phys.Rev.} {\bf D88}
  (2013)  111501},
\href{http://arxiv.org/abs/1309.7644}{{\tt arXiv:1309.7644 [hep-ph]}}.

\bibitem{Kovner:2013ona}
A.~Kovner, M.~Lublinsky, and Y.~Mulian,   
\href{http://dx.doi.org/10.1103/PhysRevD.89.061704}{{\em Phys.Rev.} {\bf D89}
  (2014) no.~6, 061704},
\href{http://arxiv.org/abs/1310.0378}{{\tt arXiv:1310.0378 [hep-ph]}}.

\bibitem{Fadin:1998py}
V.S.~Fadin and L.~Lipatov, 
\href{http://dx.doi.org/10.1016/S0370-2693(98)00473-0}{{\em
  Phys.Lett.} {\bf B429} (1998)  127--134},
\href{http://arxiv.org/abs/hep-ph/9802290}{{\tt arXiv:hep-ph/9802290}}.

\bibitem{Ciafaloni:1998gs}
M.~Ciafaloni and G.~Camici, 
\href{http://dx.doi.org/10.1016/S0370-2693(98)00551-6}{{\em
  Phys.Lett.} {\bf B430} (1998)  349--354},
\href{http://arxiv.org/abs/hep-ph/9803389}{{\tt arXiv:hep-ph/9803389}}.

\bibitem{Lipatov:1976zz}
L.~Lipatov, 
{\em Sov.J.Nucl.Phys.} {\bf 23} (1976)  338--345.

\bibitem{Kuraev:1977fs}
E.~Kuraev, L.~Lipatov, and V.S.~Fadin, 
{\em Sov.Phys.JETP} {\bf 45} (1977)  199--204.

\bibitem{Balitsky:1978ic}
I.~Balitsky and L.~Lipatov, 
{\em Sov.J.Nucl.Phys.} {\bf 28} (1978)  822--829.

\bibitem{Salam:1998tj}
G.P.~Salam,
  \href{http://dx.doi.org/10.1088/1126-6708/1998/07/019}{{\em JHEP} {\bf 9807}
  (1998)  019},
\href{http://arxiv.org/abs/hep-ph/9806482}{{\tt arXiv:hep-ph/9806482}}.

\bibitem{Ciafaloni:1999yw}
M.~Ciafaloni, D.~Colferai, and G.P.~Salam,  
\href{http://dx.doi.org/10.1103/PhysRevD.60.114036}{{\em
  Phys.Rev.} {\bf D60} (1999)  114036},
\href{http://arxiv.org/abs/hep-ph/9905566}{{\tt arXiv:hep-ph/9905566}}.

\bibitem{Ciafaloni:2003rd}
M.~Ciafaloni, D.~Colferai, G.P.~Salam, and A.M.~Stasto,
  \href{http://dx.doi.org/10.1103/PhysRevD.68.114003}{{\em Phys.Rev.} {\bf D68}
  (2003)  114003},
\href{http://arxiv.org/abs/hep-ph/0307188}{{\tt arXiv:hep-ph/0307188}}.

\bibitem{Vera:2005jt}
A.~Sabio~Vera, 
  \href{http://dx.doi.org/10.1016/j.nuclphysb.2005.06.003}{{\em Nucl.Phys.}
  {\bf B722} (2005)  65--80},
\href{http://arxiv.org/abs/hep-ph/0505128}{{\tt arXiv:hep-ph/0505128}}.

\bibitem{Triantafyllopoulos:2002nz}
D.N.~Triantafyllopoulos,  
\href{http://dx.doi.org/10.1016/S0550-3213(02)01000-3}{{\em Nucl.Phys.} {\bf
  B648} (2003)  293--316},
\href{http://arxiv.org/abs/hep-ph/0209121}{{\tt arXiv:hep-ph/0209121
  }}.

\bibitem{Avsar:2011ds}
E.~Avsar, A.M.~Stasto, D.N.~Triantafyllopoulos, and D.~Zaslavsky,
  \href{http://dx.doi.org/10.1007/JHEP10(2011)138}{{\em JHEP} {\bf 1110} (2011)
   138},
\href{http://arxiv.org/abs/1107.1252}{{\tt arXiv:1107.1252 [hep-ph]}}.

\bibitem{Lappi:2015fma}
T.~Lappi and H.~M{\"a}ntysaari, 
\href{http://arxiv.org/abs/1502.02400}{{\tt arXiv:1502.02400 [hep-ph]}}.

\bibitem{Andersson:1995ju}
B.~Andersson, G.~Gustafson, and J.~Samuelsson, 
\href{http://dx.doi.org/10.1016/0550-3213(96)00114-9}{{\em Nucl.Phys.} {\bf
  B467} (1996)  443--478}.

\bibitem{Beuf:2014uia}
G.~Beuf,  
\href{http://dx.doi.org/10.1103/PhysRevD.89.074039}{{\em
  Phys.Rev.} {\bf D89} (2014)  074039},
\href{http://arxiv.org/abs/1401.0313}{{\tt arXiv:1401.0313 [hep-ph]}}.

\bibitem{Motyka:2009gi}
L.~Motyka and A.M.~Stasto,
  \href{http://dx.doi.org/10.1103/PhysRevD.79.085016}{{\em Phys.Rev.} {\bf D79}
  (2009)  085016},
\href{http://arxiv.org/abs/0901.4949}{{\tt arXiv:0901.4949 [hep-ph]}}.

\bibitem{Mueller:1993rr}
A.H.~Mueller,
\href{http://dx.doi.org/10.1016/0550-3213(94)90116-3}{{\em Nucl.Phys.} {\bf
  B415} (1994)  373--385}.

\bibitem{Iancu:2014kga}
E.~Iancu,
  \href{http://dx.doi.org/10.1007/JHEP10(2014)095}{{\em JHEP} {\bf 1410} (2014)
   95},
\href{http://arxiv.org/abs/1403.1996}{{\tt arXiv:1403.1996 [hep-ph]}}.

\bibitem{Iancu:2003xm}
E.~Iancu and R.~Venugopalan,
\href{http://arxiv.org/abs/hep-ph/0303204}{{\tt arXiv:hep-ph/0303204}}.

\bibitem{Alvioli:2012ba}
M.~Alvioli, G.~Soyez, and D.N.~Triantafyllopoulos,   
\href{http://dx.doi.org/10.1103/PhysRevD.87.014016}{{\em Phys.Rev.}
  {\bfseries D87} (2013) 014016},
\href{http://arxiv.org/abs/1212.1656}{{\ttfamily arXiv:1212.1656 [hep-ph]}}.

\bibitem{prep}
E.~Iancu, J.D.~Madrigal, A.H.~Mueller, G.~Soyez and D.N.~Triantafyllopoulos, \emph{in preparation}.


\bibitem{Binosi:2003yf}
D.~Binosi and L.~Theussl,  
\href{http://dx.doi.org/10.1016/j.cpc.2004.05.001}{{\em
  Comput.Phys.Commun.} {\bf 161} (2004)  76--86},
\href{http://arxiv.org/abs/hep-ph/0309015}{{\tt arXiv:hep-ph/0309015
 }}.

\end{thebibliography}

\providecommand{\href}[2]{#2}\begingroup\raggedright\endgroup

\end{document}